**P o S**

PROCEEDINGS
OF SCIENCE

# Radiation Protection at ELI Beamlines: A Unique Laser Driven Accelerator Facility

**A. Cimmino,**[a,*] **D. Horváth,**[a] **V. Olšovcová,**[a] **V. Stránský,**[a] **R. Truneček,**[a] **A. Tsinganis**[a] **and R. Versaci**[a]

[a]*ELI Beamlines, Za Radnici 835, 25241 Dolni Brezany, Czech Republic*

*E-mail:* anna.cimmino@eli-beamlines.eu

The Extreme Light Infrastructure (ELI) Beamlines is a laser driven accelerator facility located in the outskirts of the city of Prague. With its state-of-the-art lasers, it will carry out an ambitious and diverse research program. Activities at ELI Beamlines can be broken down in complementary areas of scientific interest: development and testing of novel technologies for multi-PW laser systems, plasma physics, high field physics experiments, production of femtosecond secondary sources of ionizing radiation (extreme ultraviolet radiation, X rays, gamma, electrons, and protons) to be used in interdisciplinary applications in physics, biology, medicine, and material sciences. In-house experiments are already taking place since the first half of 2018, first user calls and experiments started in 2019.

In this contribution, the ELI Beamlines accelerator facility and its current status of operation are presented in more details. Particular attention is paid to the design and implementation of the radiation protection program, with emphasis on the unique challenges this laser facility poses in terms of radiation safety.

*
Speaker

## 1. ELI Beamlines

ELI Beamlines [1] is a laser driven user facility located south of Prague. Its experimental building houses four main laser systems: L1 (ALLEGRA), L2 (AMOS), L3 (HAPLS), and L4 (ATON). At the time of this contribution, L1 and L3 are in operation, L4 is in commissioning, and installation is ongoing in L2. Relevant laser parameters are listed in tab. 1. The lasers are distributed to several experimental stations, eight of which produce ionizing radiation. The HHG and PXS stations are dedicated to the production of high brilliance X-ray beams with energies between 1-30 keV. LUIS aims at the production of initially spontaneous and subsequently coherent photon radiation using electron beams (600 MeV maximum energy and < 1% peak spread). LEAP, instead, will accelerate electrons up to several GeVs. ELIMAIA has as final goal ion acceleration, but in its first phase it will accelerate protons up to 250 MeV. TERESA, is a temporary test bed for proton (10-15 MeV) and electron (100-150 MeV) acceleration. E2, instead, will generate ultra-fast and bright hard X-ray using a PW class laser with a 10 Hz repetition rate. Finally, the P3 station, with its 50 $m^3$ vacuum chamber, pursues an ambitious experimental program in high-field laser-plasma interaction as well as high-energy density physics. In-house experiments have been taking place since the first half of 2018. First user calls were issued in 2019, with first user experiments were successfully performed later that same year.

| Laser | Energy [J] | Power [TW] | Rate [Hz] |
|---|---|---|---|
| L1 (ALLEGRA) | 0.03(p) 0.1(t) | 1.5(p) 5(t) | $10^3$ |
| L2 (AMOS) | 2(t) | $10^3$(t) | 50(t) |
| L3 (HAPLS) | 30 | 333(p) $10^3$(t) | 3.3(p) 10(t) |
| L4 (ATON) | $2 \cdot 10^3$(t) | $10^4$(t) | 0.1(t) |

**Table 1:** Present (p)and target (t) laser parameters.

## 2. Radiation Protection Challenges

A laser based facility as unique as ELI Beamlines poses unique challenges from a radiation protection (RP) view point.The most noteworthy are here summarized:

- **Wide Penetrations in walls and ceilings** are present to transport lasers and to house supporting technologies, thus weakening the shielding.

- The **source term** descriptions, needed for Monte Carlo simulation, are not as well known as in conventional accelerators and are subject to high uncertainty.

- The **ultra-short laser pulses** will produce bursts of prompt radiation too short ($\sim$fs) to be reliably detected by conventional active dosimeters.

- Laser-target interactions create strong **electromagnetic pulses** potentially damaging for radiation detector electronics. Thus a combination of active and passive detectors is required.

- Experimental halls at ELI Beamlines are classified as ISO 5 to 7 (EN ISO 14644) **clean rooms**. This significantly limits available choices for shielding materials and monitoring detectors.

### 3. Radiation Protection Aspects and Methodology

The RP group at ELI Beamlines ensures that personnel, users, general public, and the environment are protected from potentially harmful effects of ionizing radiation linked to activities performed within the facility. RP assessments are driven by FLUKA Monte Carlo (MC) simulations [2]. Prompt and residual ambient dose rates, as well as induced radioactivity, are evaluated and simulation results are used to set constrains on engineering designs, set occupancy limits for users and personnel, design dumps and local shielding, plan maintenance/upgrade interventions, optimize material use, and manage radioactive waste. The civil structure of ELI Beamlines was designed so that the dose to personnel is kept < 1 mSv/year, while levels to the public are kept < 50 μSv/year.

Conjointly with MC simulations, various protection layers are in place at ELI Beamlines. These include: passive shielding, warning signs, audio/visual alarms, safety trainings, a personal safety interlock system (PSI), and radiation monitoring systems. The PSI is designed to confine hazardous area, promptly alerts to the presence of the hazard and, if possible, automatically terminates it. The radiation monitoring detectors, instead, comprise of LB 6419 detector [3] for simultaneous measurement of neutron and photon radiation, inorganic scintillators (YAG:Ce and LuAG:Ce) mounted on experimental chambers to detect electrons and photons, and Geiger-Müller counters. All these detectors are interfaced with the PSI. Hand-held contamination monitors are available for each experimental hall. Additionally to the active systems described above, optically stimulated luminescence dosimeters (OSL) of beryllium oxide, used for photon dosimetry and successfully tested in pulsed fields [4], are placed throughout the experimental building. Lithium fluoride thermoluminescent dosimeters, instead, are sensitive to thermal neutrons and/or gammas and are used in the experimental halls. This passive system has the advantage that it does not need to be protected against electromagnetic pulses. Finally, a long term environmental monitoring around the facility premises is done by sole use of OSLs.

### 4. Conclusions

The current status of the ELI Beamlines facility was presented. Emphasis was put on the RP program: Monte Carlo simulations, monitoring system, interlocks, and personal dosimetry. It was shown that same characteristics that make ELI Beamline a one of a kind center of excellence for laser-driven acceleration research create unique challenges from an RP view point.